\documentclass[12pt]{article}
\title{
\bf Ensemble of Vortex Loops in the\\ 
Abelian-Projected SU(3)-Gluodynamics}  
\author{Dmitri Antonov 
\thanks{E-mail address: 
{\tt antonov@vxitep.itep.ru}}{\,} 
\thanks {Address after October 1999: INFN-Sezione di Pisa, 
Universit\'a degli studi di Pisa, Dipartimento di Fisica, 
Via Buonarroti, 2 - Ed. B - 56127 Pisa, Italy.} 
\\
{\it Institute of Theoretical and Experimental Physics,}\\
{\it B. Cheremushkinskaya 25, RU-117 218 Moscow, Russia}} 
\date{}
\begin{document}
\maketitle
\vspace{1mm}
\centerline{\bf Abstract}
\vspace{3mm}
\noindent
Grand canonical ensemble of 
small vortex loops emerging in the London limit of the 
effective Abelian-projected theory of the $SU(3)$-gluodynamics 
is investigated in the dilute gas approximation. An essential 
difference of this system from the $SU(2)$-case is the presence of 
{\it two} interacting gases of vortex loops. Two  
alternative representations for the partition function of such a 
grand canonical ensemble are 
derived, and one of them, 
which is a representation in terms of the integrals over 
vortex loops, is employed for the evaluation of the correlators of 
both kinds of loops in the low-energy limit.

\newpage

In a recent paper~\cite{ijmpa}, the grand canonical 
ensemble of small vortex loops, existing in the Abelian Higgs model, 
have been investigated. Such loops are nothing else but the 4D 
analogue of the vortex dipoles, which are present in the usual 
Ginzburg-Landau theory. In particular, it has been demonstrated 
that the summation over the grand canonical ensemble of  
small vortex loops leads to an effective Sine-Gordon type theory of 
the massive Kalb-Ramond field~\cite{kalb} 
({\it cf.} 2D- and 3D cases studied 
in Refs.~\cite{23D, kleinert}). 
This field describes a dual vector boson, which 
therefore acquires an additional mass due to the Debye screening in the 
gas of vortex loops. Furthermore, a representation of the partition 
function of such a gas directly in terms of the integral over 
vortex loops as well as the related effective potential of those
have been discussed. 
Such a representation then turned out to be useful for 
the evaluation of the bilocal correlator of vortex loops in the 
low-energy limit. These calculations demonstrate the importance 
of treating the topological 
defects in the Abelian Higgs model 
and Ginzburg-Landau theory as forming the ensembles, 
rather than individual ones.

An interest to the study of vortex loops in the Abelian Higgs model 
is motivated by the fact that the dual Abelian Higgs model is discussed 
to be relevant to the description of confinement in the 
$SU(2)$-gluodynamics~\cite{suz}. This agreement is based on the 
method of Abelian projections~\cite{abpr} (for a recent progress 
see~\cite{digiacomo, max, reinhardt, kondo, wipf, plb, komarov, dec, 
correl2}, 
for a review see~\cite{proc}) 
and the so-called 
Abelian dominance hypothesis~\cite{abdom}, according to which 
off-diagonal (in the sense of the Cartan decomposition) degrees of freedom
are inessential for confinement and can be disregarded. In the spirit 
of this hypothesis, the following partition function describing 
an effective $[U(1)]^2$ gauge invariant 
Abelian-projected theory of the realistic 
$SU(3)$-gluodynamics has been proposed~\cite{maedan}~\footnote{
Throughout the present Letter, 
all the investigations will be performed in the Euclidean space-time.}

$$
{\cal Z}=\int {\cal D}\vec B_\mu{\cal D}\chi_a{\cal D}\chi_a^{*}
\delta\left(
\sum\limits_{a=1}^{3}\theta_a\right)\times
$$

\begin{equation}
\label{1}
\times\exp\left\{-\int d^4x\left[\frac14\vec F_{\mu\nu}^2+
\sum\limits_{a=1}^{3}\left[\left|\left(\partial_\mu-
ig_m\vec\varepsilon_a\vec B_\mu\right)\chi_a\right|^2+
\lambda\left(\left|\chi_a\right|^2-\eta^2\right)^2\right]\right]
\right\}.
\end{equation}
Here, $\vec F_{\mu\nu}=\partial_\mu\vec B_\nu-\partial_\nu\vec B_\mu$ 
stands for the field strength tensor of the magnetic 
vector potential $\vec B_\mu
\equiv\left(B_\mu^3, B_\mu^8\right)$ dual to the electric one 
$\vec A_\mu\equiv\left(A_\mu^3, A_\mu^8\right)$. Next in Eq.~(\ref{1}), 
$\chi_a=\left|\chi_a\right|{\rm e}^{i\theta_a}$, $a=1,2,3$, are 
effective Higgs fields describing condensed magnetic monopoles, whose 
magnetic charge $g_m$ is expressed via the QCD coupling constant  
$g_{\rm QCD}$ as $g_m=4\pi/g_{\rm QCD}$. Finally in Eq.~(\ref{1}), 

$$\vec\varepsilon_1=\left(1,0\right),~~ 
\vec\varepsilon_2=\left(-\frac12, -\frac{\sqrt{3}}{2}\right),~~ 
\vec\varepsilon_3=\left(-\frac12, \frac{\sqrt{3}}{2}\right)$$
stand for the so-called root vectors, which play the role of the 
structural constants in the algebra  
$\left[\vec H, E_{\pm a}\right]=\pm\vec\varepsilon_a E_{\pm a}$.   
Here, the operators 
$\vec H\equiv\left(H_1, H_2\right)=\left(T_3, T_8\right)$ generate 
the Cartan subalgebra,   
where from now on  
$T_i\equiv\frac{\lambda_i}{2}$, $i=1,\ldots, 8$, are just  
the $SU(3)$-generators. We have also introduced  
the so-called step operators $E_{\pm a}$'s (else called 
raising operators for positive $a$'s and lowering operators otherwise) 
by redefining the rest (non-diagonal) $SU(3)$-generators as follows 

$$
E_{\pm 1}=\frac{1}{\sqrt{2}}\left(T_1\pm iT_2\right),~~ 
E_{\pm 2}=\frac{1}{\sqrt{2}}\left(T_4\mp iT_5\right),~~ 
E_{\pm 3}=\frac{1}{\sqrt{2}}\left(T_6\pm iT_7\right).
$$
Clearly, these operators are non-Hermitean 
in the sense that $\left(E_a\right)^{\dag}=E_{-a}$. For bookkeeping 
purposes, it is worth listing the remaining commutation relations, 
completing the Lie algebra, which read

$$
\left[E_{\pm a}, E_{\pm b}\right]=\mp\frac{1}{\sqrt{2}}\varepsilon_{abc}
E_{\mp c}~~ {\rm and}~~ \left[E_a, E_{-b}\right]=\delta_{ab}
\vec\varepsilon_a\vec H.$$
Notice also that due to the fact that the original 
$SU(3)$ group is special, the phases of the three magnetic Higgs fields 
are not independent and 
should obey the constraint $\sum\limits_{a=1}^{3}\theta_a=0$. The latter 
one has been imposed by the introduction of the corresponding 
$\delta$-function into the functional integral on the R.H.S. of 
Eq.~(\ref{1}).

Before proceeding with the study of the model~(\ref{1}), 
it is worth mentioning its certain feature owing to which this model is 
not quite adequate to the description of confinement in the 
real $SU(3)$-gluodynamics. Its essence is that the model~(\ref{1}) 
describes only the sector of the full yet unknown 
Abelian-projected theory of the $SU(3)$-gluodynamics, where 
antimonopoles are completely absent. Clearly, the expected full  
theory should contain the antimonopole sector as well. Possible  
interference between these two sectors is up to now unclear and 
should be clarified by further investigations.

In what follows, we shall be interested in the study of the 
model~(\ref{1}) in the London limit, 
{\it i.e.}, the limit of infinitely large Higgs coupling 
constant $\lambda$. Analogously to the $SU(2)$-case, in this limit 
the model under study allows 
for an exact reformulation in terms of the integral over 
closed Abrikosov-Nielsen-Olesen type 
strings~\cite{ano}~\footnote{Notice that 
according to the lattice data~\cite{max, proc}, it is this limit of the 
Abelian-projected theories, 
in which they reveal properties similar to the real QCD.}. In this limit, 
the radial parts of the Higgs fields can be integrated out, and the 
partition function~(\ref{1}) takes the form

\begin{equation}
\label{2}
{\cal Z}=\int {\cal D}\vec B_\mu{\cal D}\theta_a
\delta\left(\sum\limits_{a=1}^{3}
\theta_a\right)
\exp\left\{-\int d^4x\left[\frac14\vec F_{\mu\nu}^2+\eta^2
\sum\limits_{a=1}^{3}\left(\partial_\mu\theta_a-g_m\vec\varepsilon_a
\vec B_\mu\right)^2\right]
\right\}.
\end{equation}
Next, the total phases $\theta_a$'s of magnetic 
Higgs fields should be decomposed 
into the singular and regular parts, $\theta_a=
\theta_a^{\rm sing}+\theta_a^{\rm reg}$~\cite{plb, komarov, dec} 
({\it cf.} also Refs.~\cite{duality, emil, correl2} 
for the $SU(2)$-case). 
Here, $\theta_a^{\rm sing}$'s describe a certain configuration of 
electric strings and are 
unambiguously related to their world-sheets $\Sigma_a$'s according to the 
equation (see the above cited Refs.)

\begin{equation}
\label{3}
\varepsilon_{\mu\nu\lambda\rho}\partial_\lambda\partial_\rho
\theta_a^{\rm sing}(x)=2\pi\Sigma_{\mu\nu}^a(x)\equiv 2\pi
\int\limits_{\Sigma_a}^{}d\sigma_{\mu\nu}\left(x^{(a)}(\xi)\right)
\delta\left(x-x^{(a)}(\xi)\right).
\end{equation}
This equation is just the covariant formulation 
of the 4D analogue of the Stokes theorem for the gradient of the 
field $\theta_a$,   
written in the local form. In Eq.~(\ref{3}), 
$x^{(a)}(\xi)
\equiv x_\mu^{(a)}(\xi)$ is a vector parametrizing the world-sheet 
$\Sigma_a$ with $\xi=\left(\xi^1, \xi^2\right)\in [0,1]\times [0,1]$ 
standing for the 
two-dimensional coordinate.

Confining and topological properties of the model~(\ref{2}) 
have been studied in Refs.~\cite{plb, komarov, dec}. This has been done 
by making use of the so-called path-integral duality 
transformation, elaborated in Refs.~\cite{kleinert, duality} 
for the usual Abelian Higgs model, 
which casts the partition function~(\ref{2}) into the following form,   

$$
{\cal Z}=\int {\cal D}x_\mu^{(a)}(\xi)\delta\left(\sum\limits_{a=1}^{3}
\Sigma_{\mu\nu}^a\right){\cal D}A_\mu^a{\cal D}h_{\mu\nu}^a\exp\left\{
-\int d^4x\left[\frac{1}{12\eta^2}\left(H_{\mu\nu\lambda}^a\right)^2+
\right.\right.$$

\begin{equation}
\label{4}
\left.\left.+\left(g_m\frac{\sqrt{3}}{2}
h_{\mu\nu}^a+\partial_\mu A_\nu^a-
\partial_\nu A_\mu^a\right)^2-i\pi\sqrt{2}h_{\mu\nu}^a\Sigma_{\mu\nu}^a
\right]\right\}
\end{equation}
with $A_\mu^a\equiv\vec\varepsilon_a \vec A_\mu$. Here, 
$H_{\mu\nu\lambda}^a=\partial_\mu h_{\nu\lambda}^a+\partial_\lambda 
h_{\mu\nu}^a+\partial_\nu h_{\lambda\mu}^a$ stands for the field 
strength tensor of the antisymmetric tensor field $h_{\mu\nu}^a$ 
(the so-called Kalb-Ramond field~\cite{kalb}). The integration over this 
field came about via some constraints resulting from the integration over 
$\theta_a^{\rm reg}$'s, whereas the integration over 
$\theta_a^{\rm sing}$'s has transformed into the integration over 
$x_\mu^{(a)}(\xi)$'s by virtue of Eq.~(\ref{3}) (Notice that since 
in what follows we shall 
be interested in effective actions rather than the integration measures, 
the Jacobian appearing during 
the change of the integration variables, $\theta_a^{\rm sing}\to 
x_\mu^{(a)}(\xi)$, 
which has been evaluated in Ref.~\cite{emil}, will not be 
discussed below and is assumed to be included into the measure ${\cal D}
x_\mu^{(a)}(\xi)$.). Also, due to this one-to-one correspondence between 
$\theta_a^{\rm sing}$'s and $\Sigma_a$'s, the 
constraint imposed by the $\delta$-function on the R.H.S. of 
Eqs.~(\ref{1}) and~(\ref{2}) has gone over into the constraint 
imposed by the 
$\delta$-function on the R.H.S. of Eq.~(\ref{4}), which relates the 
world-sheets of three types to each other, making only two of them 
really independent.

The aim of the present Letter is to treat 
Abrikosov-Nielsen-Olesen type strings in the model~(\ref{2}) in the sense 
of the grand canonical ensemble of small vortex loops, 
rather than as individual ({\it i.e.}, 
noninteracting) ones. To understand why one might expect in this case 
the appearance 
of some nontrivialities {\it w.r.t.} the Abelian-projected 
$SU(2)$-gluodynamics, let us begin with considering noninteracting 
vortex loops. This can be done by gauging the field $A_\mu^a$ away 
from Eq.~(\ref{4}) by performing the hypergauge transformation 
$h_{\mu\nu}^a\to h_{\mu\nu}^a-\frac{2}{g_m\sqrt{3}}\left(\partial_\mu 
A_\nu^a-\partial_\nu A_\mu^a\right)$ and subsequent integration 
over the Kalb-Ramond fields (see the first paper from Ref.~\cite{correl2} 
for details of this integration).
The result has the form
    
$$
{\cal Z}=\int {\cal D}x_\mu^{(a)}(\xi)\delta\left(\sum\limits_{a=1}^{3}
\Sigma_{\mu\nu}^a\right)\exp\left[-g_m\frac{\sqrt{3}}{2}\eta^3
\int d^4xd^4y 
\Sigma_{\mu\nu}^a(x)\frac{K_1(m|x-y|)}{|x-y|}
\Sigma_{\mu\nu}^a(y)\right],$$
where $m=\sqrt{3}g_m\eta$ is the mass of the fields $B_\mu^3$ and 
$B_\mu^8$, which they acquire due to the Higgs mechanism, and 
$K_1$ stands for the modified Bessel function. Finally, one of the 
three world-sheets, for concreteness $x_\mu^{(3)}(\xi)$, can be integrated 
out, which yields

$$
{\cal Z}=\int {\cal D}x_\mu^{(1)}(\xi){\cal D}x_\mu^{(2)}(\xi)\times
$$

\begin{equation}
\label{5}
\times\exp\left\{
-g_m\eta^3\sqrt{3}\int d^4xd^4y\left[\Sigma_{\mu\nu}^1(x)
\Sigma_{\mu\nu}^1(y)+\Sigma_{\mu\nu}^1(x)\Sigma_{\mu\nu}^2(y)+
\Sigma_{\mu\nu}^2(x)\Sigma_{\mu\nu}^2(y)\right]
\frac{K_1(m|x-y|)}{|x-y|}\right\}.
\end{equation}
In order to proceed from the individual strings to the grand canonical 
ensemble of interacting vortex loops, one should replace 
$\Sigma_{\mu\nu}^a(x)$, where from now on $a=1,2$, in Eq.~(\ref{5}) 
by 

$$
\Sigma_{\mu\nu}^{a{\,}{\rm gas}}(x)=\sum\limits_{k=1}^{N}n_k^{(a)}
\int d\sigma_{\mu\nu}\left(x_k^{(a)}(\xi)\right)\delta\left(
x-x_k^{(a)}(\xi)\right).
$$
Here, $n_k^{(a)}$'s stand for winding numbers, which 
we shall set to be equal $\pm 1$ ({\it cf.} Ref.~\cite{ijmpa})~\footnote{ 
This is just the essence of the dipole approximation.}. 
Performing such a replacement, one can see the crucial difference 
of the grand canonical ensemble of small vortex loops in the model 
under study from that in the Abelian-projected 
$SU(2)$-gluodynamics~\cite{ijmpa}. Namely, the system 
has now the form of two interacting gases consisting of the vortex loops 
of two kinds, while in the $SU(2)$-case the gas was built out of 
vortex loops of the only one kind. 

Analogously to that case, we shall 
treat such a grand canonical ensemble of vortex loops in the dilute 
gas approximation. According to it, characteristic sizes of loops are 
much smaller than characteristic distances between them, which in 
particular means that the vortex loops are short living objects.
Then the summation over this grand canonical ensemble 
can be most easily performed by inserting the 
unity 

\begin{equation}
\label{aux}
1=\int {\cal D}S_{\mu\nu}^a\delta\left(S_{\mu\nu}^a-
\Sigma_{\mu\nu}^{a{\,}{\rm gas}}\right)
\end{equation} 
into the R.H.S. of Eq.~(\ref{5})
(with $\Sigma_{\mu\nu}^a$ replaced by $\Sigma_{\mu\nu}^{a{\,}{\rm gas}}$) 
and representing the $\delta$-functions as the integrals over Lagrange 
multipliers. Then, the contribution of $N$ vortex loops of each kind 
to the full grand canonical ensemble takes the following form  

$$
{\cal Z}\left[\Sigma_{\mu\nu}^{a{\,}{\rm gas}}\right]
=\int {\cal D}S_{\mu\nu}^a
{\cal D}\lambda_{\mu\nu}^a\times$$

$$
\times\exp\left\{
-\left\{g_m\eta^3\sqrt{3}\int d^4xd^4y\left[S_{\mu\nu}^1(x)
S_{\mu\nu}^1(y)+S_{\mu\nu}^1(x)S_{\mu\nu}^2(y)+
S_{\mu\nu}^2(x)S_{\mu\nu}^2(y)\right]
\frac{K_1(m|x-y|)}{|x-y|}+\right.\right.$$

\begin{equation}
\label{6}
\left.\left.+i\int d^4x\lambda_{\mu\nu}^a\left(S_{\mu\nu}^a-
\Sigma_{\mu\nu}^{a{\,}{\rm gas}}\right)\right\}\right\}.
\end{equation}
After that, the desired summation is straightforward, since it technically 
parallels the one of Abelian-projected $SU(2)$-gluodynamics 
described in Ref.~\cite{ijmpa}. We have 

$$
\left\{1+\sum\limits_{N=1}^{\infty}\frac{\zeta^N}{N!}\left(
\prod\limits_{i=1}^{N}\int d^4y_i^{(1)}\int {\cal D}z_i^{(1)}(\xi)\mu
\left[z_i^{(1)}\right]\right)\times\right.$$

$$\left.\times\sum\limits_{n_k^{(1)}=\pm 1}^{}
\exp\left[i\sum\limits_{k=1}^{N}n_k^{(1)}\int d\sigma_{\mu\nu}
\left(z_k^{(1)}(\xi)\right)\lambda_{\mu\nu}^1\left(x_k^{(1)}(\xi)\right)
\right]\right\}\times$$

$$\times
\left\{ {\rm the}~ {\rm same}~ {\rm term}~ {\rm with}~ {\rm the}~ 
{\rm replacements}~ (1)\to (2)~ {\rm and}~ \lambda_{\mu\nu}^1\to 
\lambda_{\mu\nu}^2 \right\}=$$

\begin{equation}
\label{7}
=\exp\left\{2\zeta\int d^4y\left[\cos\left(\frac{\left|
\lambda_{\mu\nu}^1(y)\right|}{\Lambda^2}\right)+
\cos\left(\frac{\left|
\lambda_{\mu\nu}^2(y)\right|}{\Lambda^2}\right)\right]\right\}.
\end{equation}
Here, the world-sheet coordinate of the $k$-th vortex loop 
of the $a$-th type~\footnote{For brevity, we omit the Lorentz index.} 
$x_k^{(a)}(\xi)$ has been decomposed 
as $x_k^{(a)}(\xi)=y_k^{(a)}+z_k^{(a)}(\xi)$, where the vector 
$y_k^{(a)}\equiv\int d^2\xi x_k^{(a)}(\xi)$ describes the position 
of the vortex loop, whereas the vector $z_k^{(a)}(\xi)$ describes 
its shape. Next, on the L.H.S. of Eq.~(\ref{7}), $\mu\left[z_i^{(a)}
\right]$ stands for a certain rotation- and translation invariant 
measure of integration over the shapes of the world-sheets of the vortex 
loops, and $\zeta\propto {\rm e}^{-S_0}$ is the so-called fugacity 
(Boltzmann factor of a single vortex loop~\footnote{It is natural 
to assume that the vortex loops of different kinds have the same 
fugacity, since different $\theta_a^{\rm sing}$'s enter
the initial partition function~(\ref{2}) in the same way.}) 
of dimension $({\rm mass})^4$ 
with $S_0$ denoting the action of a single loop. In Eq.~(\ref{7}), 
we have also introduced the UV momentum cutoff $\Lambda\equiv\sqrt{
\frac{L}{a^3}}$ $\left(\gg a^{-1}\right)$, 
where $a$ is a typical size of the 
vortex loop, and $L$ is a typical distance between loops, so that 
in the dilute gas approximation under study $a\ll L$. 
Finally in 
Eq.~(\ref{7}), we have denoted $\left|\lambda_{\mu\nu}^a\right|\equiv
\sqrt{\left(\lambda_{\mu\nu}^a\right)^2}$. 
The reader is 
referred to Ref.~\cite{ijmpa} for details of a derivation 
of Eq.~(\ref{7}).
 
Note that the value of $S_0$ is approximately equal to $\sigma a^2$, 
where we have estimated the area of a vortex loop as $a^2$, and 
$\sigma$ stands for an analogue of the string tension for the 
loop, {\it i.e.}, its energy per unit area. This energy can 
be evaluated from Eq.~(\ref{5}) by virtue of the results of 
Ref.~\cite{22} and has the form 

\begin{equation}
\label{logarithm}
\sigma=2\eta^2\int d^2t\frac{K_1(|t|)}{|t|}\simeq 2\pi\eta^2
\ln\left(\frac{\lambda}{g_m^2}\right).
\end{equation}
Here, we have in the standard way~\cite{ano} set for a characteristic 
small dimensionless quantity in the model under study the value 
$\frac{g_m}{\sqrt{\lambda}}$, which is of the order of the ratio 
of $m$ to the masses of magnetic Higgs fields. Moreover, it 
has been assumed that not only $\frac{\sqrt{\lambda}}{g_m}\gg 1$, 
but also $\ln\frac{\sqrt{\lambda}}{g_m}\gg 1$, 
{\it i.e.}, the last equality on the 
R.H.S. of Eq.~(\ref{logarithm}) is valid with the logarithmic accuracy. 

Next, it is possible to integrate out the Lagrange multipliers 
by solving the saddle-point equations following from Eqs.~(\ref{6}) 
and~(\ref{7}), 

$$\frac{\lambda_{\mu\nu}^a}{\left|\lambda_{\mu\nu}^a\right|}\sin\left(
\frac{\left|\lambda_{\mu\nu}^a\right|}{\Lambda^2}\right)=
-\frac{i\Lambda^2}{2\zeta}S_{\mu\nu}^a.$$
After that, we 
arrive at the following representation for the partition function 
of the grand canonical ensemble 

$$
{\cal Z}_{\rm grand}=
\int {\cal D}S_{\mu\nu}^a
\exp\left\{-\left[
g_m\eta^3\sqrt{3}\int d^4xd^4y\left[S_{\mu\nu}^1(x)
S_{\mu\nu}^1(y)+S_{\mu\nu}^1(x)S_{\mu\nu}^2(y)+
S_{\mu\nu}^2(x)S_{\mu\nu}^2(y)\right]\times\right.\right.$$

\begin{equation}
\label{8}
\left.\left.\times
\frac{K_1(m|x-y|)}{|x-y|}+V\left[S_{\mu\nu}^1\right]+
V\left[S_{\mu\nu}^2\right]\right]\right\},
\end{equation}
which owing to Eq.~(\ref{aux}) is natural to be referred to as 
the representation in terms of the vortex loops. In Eq.~(\ref{8}), 
the effective potential of 
vortex loops reads

$$
V\left[S_{\mu\nu}^a\right]=\sum\limits_{n=-\infty}^{+\infty}
\int d^4x\left\{\Lambda^2\left|S_{\mu\nu}^a
\right|\left[\ln\left[\frac{\Lambda^2}{2\zeta}\left|S_{\mu\nu}^a\right|+
\sqrt{1+\left(\frac{\Lambda^2}{2\zeta}\left|S_{\mu\nu}^a\right|\right)^2}
\right]+2\pi i n\right]-\right.
$$

\begin{equation}
\label{9}
\left.-2\zeta
\sqrt{1+\left(\frac{\Lambda^2}{2\zeta}\left|S_{\mu\nu}^a\right|\right)^2}
\right\}.
\end{equation}

It is further instructive to illustrate the difference of such a 
partition function of two interacting gases of vortex loops from the 
case of Abelian-projected $SU(2)$-gluodynamics by studying a related 
representation in terms of a certain effective Sine-Gordon 
theory. This can be done by introducing the new integration variables 
${\cal S}_{\mu\nu}^1=\frac{\sqrt{3}}{2}\left(S_{\mu\nu}^1+S_{\mu\nu}^2
\right)$ and ${\cal S}_{\mu\nu}^2=\frac12\left(S_{\mu\nu}^1-S_{\mu\nu}^2
\right)$, which diagonalize the quadratic form in square brackets 
on the R.H.S. of Eq.~(\ref{6}). Then Eqs.~(\ref{6}) and~(\ref{7}) yield

$${\cal Z}_{\rm grand}=\int {\cal D}{\cal S}_{\mu\nu}^a
{\cal D}\lambda_{\mu\nu}^a
\exp\left\{
-g_m\eta^3\sqrt{3}\int d^4xd^4y
{\cal S}_{\mu\nu}^a(x)\frac{K_1(m|x-y|)}{|x-y|}{\cal S}_{\mu\nu}^a(y)+
\right.$$

\begin{equation}
\label{10}
\left.+2\zeta\int d^4x\left[\cos\left(\frac{\left|
\lambda_{\mu\nu}^1(x)\right|}{\Lambda^2}\right)+
\cos\left(\frac{\left|
\lambda_{\mu\nu}^2(x)\right|}{\Lambda^2}\right)\right]
-i\int d^4xh_{\mu\nu}^a{\cal S}_{\mu\nu}^a\right\},
\end{equation}
where we have denoted
$h_{\mu\nu}^1=\frac{1}{\sqrt{3}}
\left(\lambda_{\mu\nu}^1+\lambda_{\mu\nu}^2\right)$ and 
$h_{\mu\nu}^2=\lambda_{\mu\nu}^1-\lambda_{\mu\nu}^2$. The partition 
function of the desired Sine-Gordon theory can be obtained 
from Eq.~(\ref{10}) by making use of the following 
equality~\footnote{This equality can straightforwardly be proved 
by mentioning that $\partial_\mu h_{\mu\nu}^a=0$, which follows 
from the equation of motion corresponding to its L.H.S. and 
Eq.~(\ref{aux}), according to which $\partial_\mu{\cal S}_{\mu\nu}^a=0$.}
  
$$
\int {\cal D}{\cal S}_{\mu\nu}^a
\exp\left\{
-\left[g_m\eta^3\sqrt{3}\int d^4xd^4y
{\cal S}_{\mu\nu}^a(x)\frac{K_1(m|x-y|)}{|x-y|}{\cal S}_{\mu\nu}^a(y)+
i\int d^4x h_{\mu\nu}^a{\cal S}_{\mu\nu}^a\right]\right\}=
$$

$$
=\exp\left\{-\frac{1}{4\pi^2}\int d^4x\left[\frac{1}{12\eta^2}
\left(H_{\mu\nu\lambda}^a\right)^2+\frac{3g_m^2}{4}
\left(h_{\mu\nu}^a\right)^2
\right]\right\}
$$
({\it cf.} the R.H.S. with the quadratic part of the action of the 
Kalb-Ramond field on the R.H.S. of Eq.~(\ref{4}) with the field 
$A_\mu^a$ gauged away). Substituting this equality into Eq.~(\ref{10})
and performing the rescaling $\frac{h_{\mu\nu}^a}{2\pi}\to h_{\mu\nu}^a$, 
we arrive at the following representation for the partition function 
of the grand canonical ensemble of vortex loops in terms of the 
local Sine-Gordon theory, equivalent to the 
nonlocal theory~(\ref{8}), 

$${\cal Z}_{\rm grand}=\int {\cal D}h_{\mu\nu}^a\exp\left\{
-\int d^4x\left\{
\frac{1}{12\eta^2}
\left(H_{\mu\nu\lambda}^a\right)^2+\frac{3g_m^2}{4}
\left(h_{\mu\nu}^a\right)^2
-\right.\right.$$

\begin{equation}
\label{11}
\left.\left.-2\zeta\left[\cos\left(\frac{\pi}{\Lambda^2}\left|\sqrt{3}
h_{\mu\nu}^1+h_{\mu\nu}^2\right|\right)+ 
\cos\left(\frac{\pi}{\Lambda^2}\left|\sqrt{3}
h_{\mu\nu}^1-h_{\mu\nu}^2\right|\right)\right]\right\}\right\}.
\end{equation}
As we now see, an essential property of the obtained Sine-Gordon theory, 
which distinguishes it from an analogous theory describing the 
grand canonical ensemble of vortex loops in the 
Abelian-projected $SU(2)$-gluodynamics~\cite{ijmpa}, is the presence 
of two interacting Kalb-Ramond fields, while in the $SU(2)$-case there 
was only one self-interacting field. Notice that upon the expansion 
of the cosines on the R.H.S. of Eq.~(\ref{11}), 
it is straightforward to see 
that only the interaction terms of the type $\left(h_{\mu\nu}^1
\right)^{2n}\left(h_{\mu\nu}^2\right)^{2k}$ survive. In another words, 
despite of the mixing of the Kalb-Ramond 
fields in the arguments of the cosines, no terms linear in any of 
these fields appear in the action. In particular, the masses of 
both Kalb-Ramond fields, $M_1$ and $M_2$, 
can be read off from Eq.~(\ref{11}) by 
expanding the cosines up to the quadratic terms. The result reads 
$M_a^2=m^2+m_a^2\equiv Q_a^2\eta^2$, where $m_1=\frac{2\pi\eta}{\Lambda^2}
\sqrt{6\zeta}$, $m_2=\frac{2\pi\eta}{\Lambda^2}\sqrt{2\zeta}$
are the Debye masses, and we have introduced the corresponding 
magnetic charges 
$Q_1=\sqrt{3g_m^2+\frac{24\pi^2\zeta}{\Lambda^4}}$, 
$Q_2=\sqrt{3g_m^2+\frac{8\pi^2\zeta}{\Lambda^4}}$.  

Eqs.~(\ref{8}) and~(\ref{9}) can now be used for the evaluation 
of correlators of vortex loops, which due to Eq.~(\ref{aux}), 
are nothing else but the correlators of $S_{\mu\nu}^a$'s. Those are 
calculable in the low-energy limit, $\Lambda^2\left|S_{\mu\nu}^a
\right|\ll\zeta$, by considering 
the real branch of the potential~(\ref{9}), {\it i.e.}, extracting 
from the whole sum the term with $n=0$.  
The latter one 
has a simple parabolic form, and in the vicinity of its minimum 
(corresponding to the low-energy limit) 
the generating functional
for correlators of $S_{\mu\nu}^a$'s reads

$${\cal Z}\left[J_{\mu\nu}^a\right]=
\int {\cal D}{\cal S}_{\mu\nu}^a
\exp\left\{
-\left[g_m\eta^3\sqrt{3}\int d^4xd^4y
{\cal S}_{\mu\nu}^a(x)\frac{K_1(m|x-y|)}{|x-y|}{\cal S}_{\mu\nu}^a(y)+
\right.\right.
$$

$$
\left.\left.
+\frac{\Lambda^4}{2\zeta}\int d^4x\left[\frac13\left({\cal S}_{\mu\nu}^1
\right)^2+\left({\cal S}_{\mu\nu}^2\right)^2\right]+\int d^4x
\left[{\cal S}_{\mu\nu}^1\frac{J_{\mu\nu}^{+}}{\sqrt{3}}
+{\cal S}_{\mu\nu}^2J_{\mu\nu}^{-}\right]
\right]\right\},$$
where $J_{\mu\nu}^a$ is a source of $S_{\mu\nu}^a$, and 
$J_{\mu\nu}^{\pm}\equiv J_{\mu\nu}^1\pm J_{\mu\nu}^2$. Such two Gaussian 
integrals can be calculated by virtue of the following equality 

$$\int {\cal D}
{\cal S}_{\mu\nu}
\exp\left\{
-\left[g_m\eta^3\sqrt{3}\int d^4xd^4y
{\cal S}_{\mu\nu}(x)\frac{K_1(m|x-y|)}{|x-y|}{\cal S}_{\mu\nu}(y)
+\frac{\Lambda^4}{2\zeta}\int d^4x{\cal S}_{\mu\nu}^2+\int d^4x 
J_{\mu\nu}{\cal S}_{\mu\nu}\right]\right\}=$$

$$=\exp\left\{-\frac{M_2\zeta}{8\pi^2\Lambda^4}\int d^4xd^4y
J_{\mu\nu}(x)J_{\mu\nu}(y)\left(\partial_x^2-m^2\right)
\frac{K_1(M_2|x-y|)}{|x-y|}\right\},$$
and the result reads
  
$${\cal Z}\left[J_{\mu\nu}^a\right]={\cal Z}_{\rm grand}\exp\left\{
-\frac{\zeta}{8\pi^2\Lambda^4}\int d^4xd^4y\left[
M_1J_{\mu\nu}^{+}(x)J_{\mu\nu}^{+}(y)
\left(\partial_x^2-m^2\right)\frac{K_1(M_1|x-y|)}{|x-y|}+
\right.\right.$$

$$
\left.\left.+M_2J_{\mu\nu}^{-}(x)J_{\mu\nu}^{-}(y)
\left(\partial_x^2-m^2\right)\frac{K_1(M_2|x-y|)}{|x-y|}
\right]\right\}.
$$
The correlators of vortex loops following from this expression 
have the form 

$$\left<S_{\mu\nu}^1(x)S_{\lambda\rho}^1(0)\right>=
\left<S_{\mu\nu}^2(x)S_{\lambda\rho}^2(0)\right>=
\left(\delta_{\mu\lambda}\delta_{\nu\rho}-\delta_{\mu\rho}
\delta_{\nu\lambda}\right)\frac{\zeta}{2\Lambda^4}\left[
2\delta(x)-\sum\limits_{a=1}^{2}m_a^2\frac{M_a}{4\pi^2}
\frac{K_1(M_a|x|)}{|x|}\right]$$
and

$$
\left<S_{\mu\nu}^1(x)S_{\lambda\rho}^2(0)\right>=
\left(\delta_{\mu\lambda}\delta_{\nu\rho}-\delta_{\mu\rho}
\delta_{\nu\lambda}\right)\frac{\zeta m_2^2}{2\Lambda^4}
\left[\frac{M_2}{4\pi^2}\frac{K_1(M_2|x|)}{|x|}-
3\frac{M_1}{4\pi^2}\frac{K_1(M_1|x|)}{|x|}\right].$$
At this point, it is worth recalling that the original theory~(\ref{1})  
is an effective theory at large distances~\cite{maedan}, where the 
asymptotic behaviours of the obtained correlators read

$$\left<S_{\mu\nu}^1(x)S_{\lambda\rho}^1(0)\right>=
\left<S_{\mu\nu}^2(x)S_{\lambda\rho}^2(0)\right>\longrightarrow
-\left<S_{\mu\nu}^1(x)S_{\lambda\rho}^2(0)\right>\longrightarrow$$

$$
\longrightarrow 
-\left(\delta_{\mu\lambda}\delta_{\nu\rho}-\delta_{\mu\rho}
\delta_{\nu\lambda}\right)\sqrt{\frac{\pi}{2}}\left(
\frac{\eta\zeta}{\Lambda^4}\right)^2\frac{\sqrt{M_2}}{|x|^{\frac32}}
{\rm e}^{-M_2|x|}.$$
This result illustrates how the vortex loops in the grand canonical 
ensemble under study are correlated to each other. Namely, their 
correlators decrease according to the Yukawa type law with the screening 
provided by the lightest of the two full masses, $M_2$.

In conclusion, we have demonstrated that the grand canonical 
ensemble of vortex loops in the 
effective Abelian-projected theory of the $SU(3)$-gluodynamics (being 
treated in the dilute gas approximation)  
exhibits an essential property distinguishing it 
from the one of the Abelian-projected $SU(2)$-gluodynamics.
Namely, it consists of two interacting subsystems, corresponding to 
two independent types of strings, which emerge after the 
Abelian projection. An average over the shapes of the vortex loops 
with the most general rotation- and translation invariant integration 
measure leads to two
alternative field-theoretical representations of such a grand 
canonical ensemble. First of them is a representation in terms 
of an effective Sine-Gordon theory of two interacting Kalb-Ramond 
fields~(\ref{11}). It yields the (positive)
contribution to the masses $M_a$'s 
of the Kalb-Ramond fields coming about 
from the Debye screening as well as to the magnetic 
charges $Q_a$'s of these fields. The other 
representation given by Eqs.~(\ref{8}) and~(\ref{9}), which is the one 
in terms of the integral over the vortex loops, is useful for the 
evaluation of their correlators. While such a calculation is difficult 
to perform exactly due to the complicated form of the effective  
potential of the vortex loops~(\ref{9}), it turns out to be possible 
to perform it 
in the low-energy limit within an additional approximation when only the 
real branch of the potential is taken into account. As a result, the 
correlators of vortex loops have a Yukawa type asymptotic behaviours  
at large distances with the screening governed by the lightest of the 
two full masses of the Kalb-Ramond fields.

It now looks reasonable to find field-theoretical representations 
for the grand canonical ensembles of vortex loops emerging after 
the Abelian projection in the general $SU(N)$, $N>2$, case. 
In that case, there 
appear $\frac{N(N-1)}{2}-1$ independent strings. Indeed, this is just 
the number 
of possibilities for the eigenvalues of a certain (adjointly transformed) 
operator, to be diagonalized during the Abelian projection, to coincide, 
minus one constraint imposed by the $\delta$-function on the R.H.S. 
of Eq.~(\ref{1}) (with the sum over $a$ going from $1$ to $N$).
It will therefore emerge just this amount of interacting gases of 
vortex loops, leading to different Debye masses. The study of this 
system as well as its large-$N$ limit 
will be the topic of a separate publication.

\section*{Acknowledgments}
The author is indebted to Prof. Yu.A. Simonov for critical reading 
the manuscript and useful discussions, 
Prof. D. Ebert for useful discussions and 
correspondence, and    
Profs. H.G. Dosch and M.G. Schmidt
for valuable discussions.

\end{document}